\documentclass[amsmath,amssymb]{revtex4}

\usepackage{graphicx}
\usepackage{dcolumn}
\usepackage{bm}
\begin{document}
\title{On The Structure of $A=3$ Nuclei}
\author{Syed Afsar Abbas and Shakeb Ahmad}
\email{drafsarabbas@yahoo.in}
\affiliation{Department of Physics, Aligarh Muslim University, Aligarh-202 002, India.
}

\begin{abstract}
The hole in the charge distribution of $^3{\text He}$ is a major problem in 
$A=3$ nuclei. The canonical wavefucntion of $A=3$ nuclei which does well for 
electromagnetic properties of $A=3$ nuclei fails to produce the hole in $A=3$
nuclei. The hole is normally assumed to arise from explicit quark degree of
freedom. Very often quark degrees of freedom are imposed to propose a different
short range part of the wavefunction for $A=3$ to explain the hole in $^3{\text He}$. 
So an hybrid model with nucleonic degree of freedom in outer part
and quark degrees of freedom in the inner part of the nucleus have been 
invoked to understand the above problem. Here we present a different picture
with a new wavefunction working at short range within nucleonic 
degrees of freedom itself. So the above problem is explained here based entirely
on the nucleonic degree of freedom only.
\end{abstract}

\maketitle 
Clusters are playing important role in current nuclear physics. Of these
$\alpha$-clusters are known to play a prominent role. These are known to be
spontaneously emitted from heavy nuclei. Heavier clusters radioactivity was 
predicted by Sandulescu, Poenaru and Greiner in 1980~\cite{san80}.
This was subsequently confirmed a couple of years later and several heavy 
clusters like $^{14}\text{C}$, $^{24}\text{Ne}$, $^{28}\text{Mg}$, 
$^{32}\text{Si}$ etc. have been observed and well studied~\cite{po11}.

Other clustering in particular $A=3$ clustering is also possible and this was
recently studied in detail by us~\cite{afs11}. Strong experimental evidences of 
clustering of $A=3$ nuclei- $^{3}\mbox{He}$(Helion) and $^{3}\mbox{H}$(Triton)
were pointed out and a consistent theoretical understanding was attempted therein.
One would have thought that a nuclei as simple as $A=3$ should have been well
understood by now. But in keeping with complexity in the structure of another 
three body system, that of baryons
made up of three quarks, things are not all that easy.

To understand the clustering of $A=3$ nuclei we have to see what aspects are the
ones which contribute to its clustering property. Hence a better and consistent understanding
of the structure of $A=3$ is required.

To understand as to what makes $A=3$ so special let us study the 
properties of the ground state magnetic
moment of $^{3}\mbox{He}$ and $^{3}\mbox{H}$, and the $^{3}\mbox{He}$ charge density.

The degrees of freedom relevant for $^{3}\mbox{He}$ and $^{3}\mbox{H}$ are spin, isospin
and orbital space. The total wavefunction should be antisymmetric. For three nucleons the
spin wave function in standard notation are of mixed symmetry $\chi_\rho$ and $\chi_\lambda$
where $\rho$ and $\lambda$ correspond to mixed symmetric state with the first two spins in 
antisymmetric and symmetric states respectively. The corresponding mixed symmetric state
for the three nucleons are $\phi_\rho$ and $\phi_\lambda$. If we ensure full antisymmetry in
spin-isospin space then
\begin{equation}
\Psi_{\text A}=\frac{1}{\sqrt{2}}\left(\chi_\lambda\phi_\rho-\chi_\rho\phi_\lambda\right)
\end{equation}
then the orbital part of the wavefunction for $A=3$ should be completely symmetric
which we take as Fadeev wavefunction~\cite{jl81}
\begin{equation}
\Psi_{\text{A=3}}=\frac{\alpha^3}{\pi^{3/2}}e^{-\alpha^2\left(\rho^2+\chi^2\right)/2}
\end{equation}
where
\begin{eqnarray}
{\vec\rho}&=&\left({\vec r_1}-{\vec r_2}\right)/\sqrt{2}\nonumber\\
{\vec\lambda}&=&\left({\vec r_1}+{\vec r_2}-2{\vec r_3}\right)/\sqrt{6}
\end{eqnarray}
So the three nucleons are in the ground state are in orbital symmetric S-state.
Pauli Principle require that in $A=3$, the two like nucleons would be in spin
zero state. Then the magnetic dipole moment of
$^{3}\mbox{H}$ would be entirely due to the odd proton and also would arise from the
odd neutron in $^{3}\mbox{He}$.

Therefore this picture predicts
\begin{eqnarray}
\mu\left(^{3}\mbox{H}\right)&=&\mu\left(p\right)=2.79~{\text{nm}}\nonumber\\
\mu\left(^{3}\mbox{He}\right)&=&\mu\left(n\right)=-1.91~{\text{nm}}
\end{eqnarray}

This is in good agreement with the experimental value of 2.97 nm and -2.12 nm,
respectively. Thus clearly a symmetric space wavefunction in S-state does reasonably
well for the magnetic moment of $A=3$ nuclei.

What will such a wavefunction predict for the charge distribution for
$^{3}\mbox{He}$ for example. Using the symmetric space wavefunction (2)
the proton nuclear charge distribution are given in Fig. 1.
\begin{figure}
\includegraphics{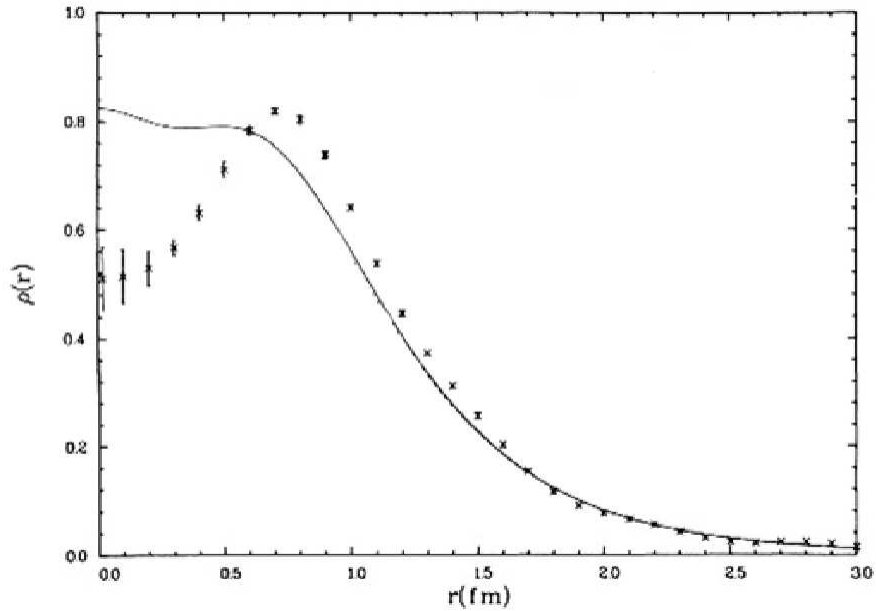}
\caption{Point nucleon $^{3}\text{He}$ theoretical charge density
and the corresponding experimental density indicated
as data points$^4$.}
\end{figure}
This is plotted against experimental results~\cite{jl81}. Such a wavefunction does 
reasonably well for $r\ge 1$ but fails miserably for small r. Experimentally
there is a 'hole' in the center of $^{3}\mbox{He}$ charge distribution.
So, the symmetric space part of the wavefunction which does do well for the
magnetic dipole moment of $A=3$ nuclei is failing badly at small distances of the
charge distribution. How to accommodate this?

One technique is to split the wavefunction in two parts. For some $r_0$ for
$r>r_0$ the wavefunction is assumed to look like that of (1) and (2)
while for $r<r_0$ it is assumed that quark degree of freedom may play
a role~\cite{hoo,afs86}. At small relative distances the nucleons are likely to
overlap strongly and as such probability of existence of 6 quark~\cite{hoo}
clusters or 9 quark~\cite{afs86} clusters towards the centre of $^{3}\mbox{He}$
would be non-negligible and thus the hole may be explained as a consequence
of these quark degrees of freedom in $A=3$ nuclei (note $^{3}\mbox{H}$ also
has a similar distribution with a hole at the centre~\cite{afs86}).

In this way the short range part of the wavefunction is arising due to the explicit 
quark degrees of freedom. In this paper we ask the question- Is it possible to
understand this hole in $A=3$ nuclei within only the nucleonic degrees of
freedom without invoking the quark degrees of freedom. 
Below we show that indeed this is possible!

We accept that there are two parts of the $A=3$ wavefunction. For some $r_0$
, for $r>r_0$ let the wavefunction be described in the canonical way through equations
(1) and (2). This is good as it explains the $A=3$ magnetic dipole moment
satisfactorily.

For nucleons in $A=3$ the degree of freedom are spin-isospin and orbital part.
In equation (1) the spin-isospin part was antisymmetric and thus the space part
was symmetric (2).

Note that it is possible to have an entirely symmetric spin-isospin wavefunction
for the $A=3$ system
\begin{equation}
\Psi_{\text S}=\frac{1}{\sqrt{2}}\left(\chi_\lambda\phi_\lambda+\chi_\rho\phi_\rho\right)
\end{equation}
If this holds then what we require is antisymmetric space wavefunction. So the question is-
is it possible to have an antisymmetric part of space wavefunction for the
ground state of $A=3$ nuclei?

We take the cue from quark model of hadrons. As is well known~\cite{jjj69} nucleons are made
up of three constituent quarks. It was found that in space-isospin-orbital space
a symmetric wavefunction worked well. Normally one would have a symmetric $L=0$
state for the ground state and in which case it was a puzzle as to how the three
spin-1/2 quarks requires symmetric spin-isospin-orbital wavefunction. One may fix for this by 
involving an additional color degree of freedom to carry the antisymmetry.
With hindsight, this was the correct way to solve the problem. 
But in the early days of quark model, without color degree of freedom, one possibility
was to seek for an $L=0$ orbital antisymmetric wavefunction. And indeed it is possible to construct $L=0$
wavefunction for baryons which is totally antisymmetric~\cite{jjj69}.
Though in keeping with Pauli principle this was possible, a difficulty was pointed
out by Mitra and Majumdar~\cite{mit66}. What they showed was that with such an 
antisymmetric orbital wavefunction for 3 quarks in $L=0$ state there would
be zeros in the form factors of proton, and such a situation does not arise experimentally. So 
this ruled out the possibility of absorbing antisymmetry of 3 quarks in the orbital
space in quark model~\cite{jjj69,mit66,kre67}. And as we know in quark model- this problem was
solved through the correct invocation of a new three color degrees of freedom.

Here for us the $L=0$ antisymmetric wavefunction of three fermions is significant.
If for $A=3$ the spin-isospin wavefunction is symmetric as in Eq.(5) then the 
orbital part should be antisymmetric here. What failed for 3 quarks in baryons
works wonderfully for 3 nucleons in $A=3$ nuclei. We see this below.

Let the completely antisymmetric wavefunction in orbital space be given as 
$f\left({\vec r_1},{\vec r_2},{\vec r_3}\right)$~\cite{jjj69}[p.31]
(the exact form of $f$ is not important for our discussion here).
Then the charge density is
\begin{equation}
\rho({\vec r})=\int d^3{\vec r_2}~|f\left({\vec r},{\vec r_2},
-\left({\vec r}+{\vec r_2}\right)\right)|^2
\end{equation}
We choose the coordinate of the three nucleons in the center of mass system in
which ${\vec r_1}+{\vec r_2}+{\vec r_3}=0$. As $f$ is antisymmetric the
integrand in Eq.(6) vanishes and thus the charge density at the origin is zero.

Thus the existence of antisymmetric orbital wavefunction for $A=3$ demands that
the charge density vanishes at $r=0$ for these nuclei. So the wavefunction 
predicts a hole at the centre of $A=3$ nuclei exactly as found for $^{3}\text{He}$ (Fig.1). 
One may expect that the finite size of nucleons would smear out the hole 
somewhat but should still leave a prominent depression in the density
near the center which would be seen as a hole (see Fig.1).

Thus we suggest that for $r<r_0$ (where $r_0\sim$1fm) the wavefunction
is $\Psi_{\text{Symmetric}}\left(\text{spin-isospin}\right)\times\Psi_{\text{Antisymmetric}}
\left(\text{orbital}\right)$ and thus the hole is naturally predicted by such a wavefunction.

Note the $\langle\Psi_{\text A}|\Psi_{\text S}\rangle =0$ (from Esq. (1) and (5))
and thus the two parts of the
wavefunction, one for $r>r_0$ (for some suitable values say $r_0\sim$1fm for
$^{3}\text{He}$) and the other $r<r_0$ with Eq.(5) would work well for
both the magnetic moment etc. and for reproducing charge/matter densities
of $A=3$ nuclei, and thus is entirely within the nucleonic degree of freedom only.

Hence both antisymmetric and symmetric orbital parts play a complementary
role in predicting the magnetic moment as well as the hole in $A=3$nuclei.
It is like two different phases (with orbital part antisymmetric in one and
completely different that is symmetric, in the other one)
coexisting for the same nucleus $A=3$.

Note that light nuclei $A=1$ and $A=2$ are all pure 'surface' nuclei. A genuine nucleus
should have a distinct 'interior' plus a clear 'surface' outside. $A=3$ is hence, the first genuine nucleus. It has managed
to build an "interior" while retaining the "surface". And the new structure
suggested here plays a fundamental role in creating a first clear cut "interior"
(which appears as a hole) while still retaining the "surface" outside.

\end{document}